\documentclass[showpacs,showkeys,11pt,
preprint,preprintnumbers,nofootinbib,
groupedaddress,superscriptaddress,amsmath,amssymb]{revtex4}
\usepackage[dvipdfm]{graphicx}
\usepackage[hypertex]{hyperref}

\begin{document}
\title{Indirect bounds on heavy scalar masses of the two-Higgs-doublet model \\
in light of recent Higgs boson searches. }
\preprint{KEK-TH-1480, UT-HET 057}
\author{Shinya~Kanemura}
\email{kanemu@sci.u-toyama.ac.jp}
\affiliation{Department of Physics, The University of Toyama, 3190 Gofuku, Toyama 930-8555, Japan}
\author{Yasuhiro~Okada}
\email{yasuhiro.okada@kek.jp}
\affiliation{KEK Theory Center, Institute of Particle and Nuclear Studies,
KEK, 1-1 Oho, Tsukuba, Ibaraki 305-0801, Japan}
\affiliation{Department of Particle and Nuclear Physics, The Graduate
University for Advanced Studies (Sokendai), 1-1 Oho, Tsukuba, Ibaraki 305-0801, Japan}
\author{Hiroyuki~Taniguchi}
\email{taniguchi@jodo.sci.u-toyama.ac.jp}
\affiliation{Department of Physics, The University of Toyama, 3190 Gofuku, Toyama 930-8555, Japan}
\author{Koji Tsumura}
\email{ko2@phys.ntu.edu.tw }
\affiliation{Department of Physics and  Center for Theoretical Sciences, National Taiwan University, Taipei, Taiwan}

\begin{abstract}
We study an upper bound on masses of additional scalar bosons 
from the electroweak precision data and theoretical constraints 
such as perturbative unitarity and vacuum stability in the two Higgs doublet model 
taking account of recent Higgs boson search results. 
If the mass of the Standard-Model-like Higgs boson is rather heavy and is
outside the allowed region by the electroweak precision data,
such a discrepancy should be compensated by contributions from the additional scalar bosons. 
We show the upper bound on masses of the additional scalar bosons
to be about $2$ $(1)$ TeV for the mass of the Standard-Model-like
Higgs boson to be $240$ $(500)$ GeV. 
\end{abstract}
\maketitle

The Standard Model (SM) for elementary particles based on the  
$SU(3)_C\times SU(2)_L \times U(1)_Y$ gauge group has been tested 
accurately\cite{Ref:PDG}. 
However, no Higgs boson has been confirmed yet. 
Discovery of the Higgs boson is the most important
issue at the Fermilab Tevatron and the CERN Large Hadron Collider (LHC). 
%
Direct searches for the Higgs boson at CERN LEP have set a lower mass bound 
on the SM Higgs boson to be $114.4$ GeV\cite{Ref:LEP-h}. 
The Tevatron experiment has excluded the mass of the SM Higgs boson 
around $160$ GeV\cite{Ref:Tevatron-h}. 
Recently, the first results from the ATLAS and CMS experiments at the
LHC have been reported\cite{Ref:atlas-h,Ref:lhc-h}. 
The Higgs boson mass around $160$ GeV and $300$--$450$ GeV has been excluded 
by the data with the integrated luminosity of about $1$ fb${}^{-1}$. 

It is well known that an upper bound on the mass of the Higgs boson is obtained by the tree
level unitarity for elastic scattering processes of longitudinally-polarized
vector bosons, such as $W_L^+W_L^-\to W_L^+W_L^-$. In the SM,  
since the scattering amplitudes are proportional to the Higgs boson mass 
in the high energy limit, a large Higgs boson mass leads to a violation
of the unitarity. Consequently the upper bound is obtained on the mass
as about $710$ GeV\cite{Ref:LQT,Ref:DW}. 
On the other hand, if the Higgs boson is absent, the scattering
amplitudes grow for high energies.  
The violation of the tree level unitarity then occurs at $\sqrt{s} \sim
1.2$ TeV, where $\sqrt{s}$ is the centre-of-mass energy of the $WW$ scattering\cite{Ref:WW}.
The LHC Higgs search experiment is expected to cover the entire range of the SM Higgs 
boson mass.  Even if the Higgs boson is not found, some new physics beyond the SM
must show up below the TeV scale. 
%
If we introduce the cutoff scale $\Lambda$ into the model, more 
sensitive upper and lower bounds are obtained on the SM Higgs boson
mass as a function of $\Lambda$\cite{Ref:Triv,Ref:HR,Ref:KM}. 

From the electroweak precision data with the theoretical study for
radiative corrections\cite{Ref:PeskinTakeuchi}, the mass of the Higgs
boson in the SM is indicated to be
$m_h = 90^{+27}_{-22}$ GeV and $m_h < 161$ GeV at the $95 \%$ Confidence
Level (CL)\cite{Ref:LEP-EW}.  
%
Notice that this indirect bound on the mass cannot be applied if new physics exists
below the TeV scale and affects the calculation for the radiative correction. 
In such a case, even if the Higgs boson is heavy,
the electroweak precision data can be satisfied by
the contribution from the new physics.

In the two Higgs doublet model (THDM), radiative corrections
to the electroweak observables have already been calculated,
and the possible allowed regions for the parameter space are
evaluated under the electroweak precision data\cite{Ref:ST-2hdm,STUinTHDM} 
and theoretical constraints\cite{Ref:KKT,Ref:AAN,Ref:DM,Ref:NS,Ref:KKO}. 
Flavor physics data such as $b\to s\gamma$\cite{Ref:Barger,bsg}, 
$B\to \tau\nu$\cite{btaunu} and tau leptonic decays\cite{tauleptonicdecay,tauleptonicdecay2}
in the THDM can further constrain the parameter space depending on types of Yukawa interactions. 
In particular, the mass of charged Higgs bosons is bounded from the $b \to s \gamma$ data
to be greater than $295$ GeV\cite{bsg2} by assuming the Type-II Yukawa interaction. 

In this Letter, in light of recent Higgs boson searches, 
we reanalyze the constraint on the parameters in the THDM
by using the electroweak precision data and the theoretical constraints
from tree level unitarity and vacuum stability. 
In particular, we show an upper bound on the masses of the 
additional heavy scalar bosons can be obtained in the THDM 
depending on the mass of the SM-like Higgs boson. 
For a relatively large mass of the SM-like Higgs boson, a 
relatively large mass difference between the CP-odd Higgs boson
and the charged Higgs boson is required in order to satisfy
bounds from the electroweak precision data. They are bounded
from above by the theoretical constraints.
For a SM Higgs boson mass to be $240$ $(500)$ GeV, an upper bound 
on the mass of the CP-odd Higgs boson of about $2$ $(1)$ TeV is obtained. 

The most general THDM with the doublet fields $\Phi_1$ and $\Phi_2$ 
are constrained by flavor changing neutral current (FCNC) processes. 
We here consider the model with the softly-broken discrete $Z_2$ symmetry
under $\Phi_1 \to \Phi_1$ and $\Phi_2 \to - \Phi_2$ to avoid FCNC constraints\cite{Ref:GW}.
There are four kind of Yukawa interaction under the discrete symmetry\cite{Ref:Barger}.
In this Letter, we do not specify the type of the Yukawa interaction
because it does not affect the following discussions.   	
The Higgs potential is then given by 
\begin{align}
{\mathcal V}^\text{THDM}
&= +m_1^2\Phi_1^\dag\Phi_1+m_2^2\Phi_2^\dag\Phi_2
-m_3^2\left(\Phi_1^\dag\Phi_2+\Phi_2^\dag\Phi_1\right)
+\frac{\lambda_1}2(\Phi_1^\dag\Phi_1)^2
+\frac{\lambda_2}2(\Phi_2^\dag\Phi_2)^2\nonumber \\
&\qquad+\lambda_3(\Phi_1^\dag\Phi_1)(\Phi_2^\dag\Phi_2)
+\lambda_4(\Phi_1^\dag\Phi_2)(\Phi_2^\dag\Phi_1)
+\frac{\lambda_5}2\left[(\Phi_1^\dag\Phi_2)^2
+(\Phi_2^\dag\Phi_1)^2\right]. \label{Eq:HiggsPot}
\end{align}
The soft-breaking mass parameter $m_3^2$ and the coupling constant
$\lambda_5$ are complex in general. We here take them to be real
assuming that CP is conserved in the Higgs sector.  

The Higgs doublets $\Phi_i(i=1,2)$ can be written in terms of
the component fields as 
\begin{align}
\Phi_i=\begin{pmatrix}i\,\omega_i^+\\\frac1{\sqrt2}(v_i+h_i-i\,z_i)
\end{pmatrix},
\end{align}
where the vacuum expectation values (VEVs) $v_1$ and $v_2$ satisfy $\sqrt{v_1^2+v_2^2}=v \simeq 246$
GeV. The mass eigenstates are obtained by rotating the component fields as  
\begin{align}
\begin{pmatrix}h_1\\h_2\end{pmatrix}=\text{R}(\alpha)
\begin{pmatrix}H\\h\end{pmatrix},\quad
\begin{pmatrix}z_1\\z_2\end{pmatrix}=\text{R}(\beta)
\begin{pmatrix}z\\A\end{pmatrix},\quad
\begin{pmatrix}\omega_1^+\\\omega_2^+\end{pmatrix}=\text{R}(\beta)
\begin{pmatrix}\omega^+\\H^+\end{pmatrix},
\end{align}
where $\omega^\pm$ and $z$ are the Nambu-Goldstone bosons, $h$, $H$, 
$A$ and $H^\pm$ are respectively two CP-even, one CP-odd and charged
Higgs bosons, and  
\begin{align}
\text{R}(\theta)=\begin{pmatrix}\cos\theta&-\sin\theta\\
\sin\theta&\cos\theta\end{pmatrix}. 
\end{align}
The eight parameters $m_1^2$--$m_3^2$ and
$\lambda_1$--$\lambda_5$ are replaced by the VEV $v$, the mixing angles
$\alpha$ and $\beta(=\tan^{-1}\frac{v_2}{v_1})$, the Higgs boson masses
$m_h^{},m_H^{},m_A^{}$ and $m_{H^\pm}^{}$, and the soft $Z_2$ breaking parameter
$M^2=m_3^2/(\cos\beta\sin\beta)$. 
In particular, the quartic coupling constants are expressed
in terms of physical Higgs boson masses, mixing angles and
the soft $Z_2$ breaking mass parameter $M^2$ as 
\begin{align}
\lambda_1 &= \frac1{v^2\cos^2\beta}\left(-M^2\sin^2\beta
+m_H^2\cos^2\alpha+m_h^2\sin^2\alpha\right),\label{Eq:lam1}\\
\lambda_2 &= \frac1{v^2\sin^2\beta}\left(-M^2\cos^2\beta
+m_H^2\sin^2\alpha+m_h^2\cos^2\alpha\right),\label{Eq:lam2}\\
\lambda_3 &= \frac1{v^2}\left[-M^2
+(m_H^2-m_h^2)\frac{\sin2\alpha}{\sin2\beta}+2m_{H^+}^2\right]\label{Eq:lam3},\\
\lambda_4 &= \frac1{v^2}(M^2+m_A^2-2m_{H^+}^2),\label{Eq:lam4}\\
\lambda_5 &= \frac1{v^2}(M^2-m_A^2).\label{Eq:lam5} 
\end{align}
The coupling constants of the CP-even Higgs bosons with the weak boson
$h WW$ and $H WW$ are proportional to
$\sin(\beta-\alpha)$ and $\cos(\beta-\alpha)$.
When $\sin(\beta-\alpha) =1$, only $h$ couples to the gauge bosons
and behaves as the SM Higgs boson. We call this limit as the SM-like limit. 

As discussed in Ref.~\cite{kosy}, the masses of the heavy Higgs bosons 
($H$, $A$ and $H^\pm$) are expressed for $M \gtrsim v$ by 
\begin{align}
m_{\Phi}^2 \sim M^2 + \lambda_i v^2 [+ {\mathcal O}(\lambda v^2/M^2)],   
\end{align}
while the mass of $h$ is the SM-like form $\sim \lambda_i v^2$.  
When $M^2 \gg \lambda_i v^2$ the heavier Higgs bosons have the common mass
$\sim M$. In this case, the effect of these bosons decouples in the
large mass limit and the low energy theory becomes the SM with
$h$ being at the electroweak scale as the SM Higgs boson.
On the contrary, when $M^2 \sim \lambda_i v^2$ the effect of these bosons
does not decouple, 
and so-called nondecoupling effects appear in the low energy observables. 
Notice that the mass difference between the heavy Higgs bosons is 
independent of $M$, so that the effect on the low energy observables
can be large if the mass differences are not small.
We also note that the mass difference between heavy Higgs bosons
are related to the violation of the custodial $SU(2)$
symmetry\cite{pomarol},
which causes significant deviation in the electroweak rho parameter from the
SM prediction in the positive direction.
As we see soon below, this positive contribution to the rho parameter
(or the $T$ parameter\cite{Ref:PeskinTakeuchi}) can be used to compensate the negative contribution
of the heavy SM-like Higgs boson.

New physics effects on the electroweak oblique parameters are 
parameterized by the $S$, $T$ and $U$ parameters\cite{Ref:PeskinTakeuchi}.   
By fixing $U=0$, the central values of $S$ and $T$ are given by\cite{Ref:PDG} 
\begin{align}
S=0.03\pm0.09,\, T=0.07\pm0.08,\, (\rho_{ST}^{}=0.87),
\end{align} 
where $\rho_{ST}^{}$ is the correlation parameters for the $\chi^2$
analysis. The origin $S=T=0$ corresponds to the SM prediction 
for the reference value $m_h=117$ GeV.
The other SM parameters are chosen as $\widehat{s_Z^{}}^2 = 0.23124 \pm 0.00016, 
\alpha_S^{} = 0.01183 \pm 0.0016, m_t = 173 \pm 1.3$ GeV and 
$G_F = 1.16639\times 10^{-5}$ GeV$^{-2}$.

In the THDM, the contributions to the electroweak parameters from 
the {\it scalar boson loops} are given by\cite{Ref:ST-2hdm}
\begin{align}
S_\Phi^{} =& -\frac1{4\pi} \bigl[ F'_\Delta(m_{H^\pm}^{},m_{H^\pm}^{})
-\sin^2(\beta-\alpha) F'_\Delta(m_H^{},m_A^{}) 
-\cos^2(\beta-\alpha)F'_\Delta(m_h^{},m_A^{})
\bigr],\\
T_\Phi^{} =& -\frac{\sqrt2G_F}{16\pi^2\alpha_\text{EM}^{}}
\Bigl\{ 
-F_\Delta(m_A^{},m_{H^+}^{}) 
+\sin^2(\beta-\alpha) \bigl[ F_\Delta(m_H^{},m_A^{})
-F_\Delta(m_H^{},m_{H^+}^{}) \bigr] \nonumber \\
&\hspace{15ex}+\cos^2(\beta-\alpha) \bigl[ F_\Delta(m_h^{},m_A^{})
-F_\Delta(m_h^{},m_{H^+}^{}) \bigr] \Bigr\},
\end{align}
where 
\begin{align}
F_\Delta(m_0,m_1) &= F_\Delta(m_1,m_0) 
= \frac{m_0^2+m_1^2}2 -\frac{m_0^2m_1^2}{m_0^2-m_1^2}\ln\frac{m_0^2}{m_1^2},\\
{F_\Delta}'(m_0,m_1) &={F_\Delta}'(m_1,m_0) 
= -\frac13 \Bigl[ \frac43
-\frac{m_0^2 \ln m_0^2 -m_1^2 \ln m_1^2}{m_0^2-m_1^2} 
-\frac{m_0^2+m_1^2}{(m_0^2-m_1^2)^2}F_\Delta(m_0,m_1) \Bigr]. 
\end{align}
For the case with $m_0\approx m_1$, we have  
\begin{align}
F_\Delta(m_0,m_1) &\approx 
\frac{2(m_0-m_1)^2}3 -\frac{(m_0-m_1)^4}{30m_1^3} +\cdots,\\
{F_\Delta}'(m_0,m_1) &\approx 
+\frac13 \ln{m_1^2} +\frac{m_0-m_1}{6m_1} +\cdots.
\end{align}
When all the additional heavy scalar bosons are degenerate
$m_A^{} = m_H^{} = m_{H^\pm}^{}$, we obtain $S_\Phi=T_\Phi=0$. 
In the SM-like limit $\sin(\beta-\alpha)=1$ with the further assumption $m_H^{}=m_A^{}$,  
we have 
\begin{align}
S_\Phi &\approx -\frac1{12\pi} \ln \frac{m_{H^\pm}^2}{m_A^2},\\
T_\Phi &\approx +\frac{\sqrt2G_F}{12\pi^2\alpha_\text{EM}^{}}
(m_A^{}-m_{H^\pm}^{})^2.\label{Eq:F5}
\end{align}

\begin{figure}[tb]
\centering
\includegraphics[height=6cm]{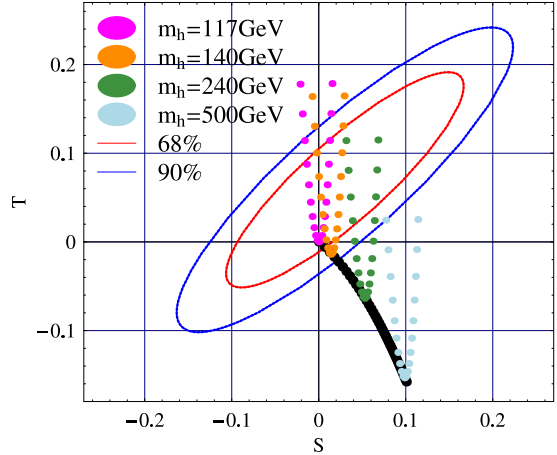}
\caption{The $\chi^2$ analysis in the ($S,T$) plane is shown in the THDM 
where the SM-like Higgs boson is taken to be $117, 140, 240$ and $500$ GeV, 
with the SM-like limit $\sin(\beta-\alpha)=1$ and $m_{H^\pm}^{}=300$ GeV.
The mass of heavy neutral Higgs bosons  $m_A^{} = m_H^{}$ is varied from 
$200$ GeV to $400$ GeV by the $10$ GeV step (dots: from left to right). 
Ellipses correspond to electroweak precision limits with 
$68\%\, (\sqrt{2.30} \sigma)$ and $90\%\, (\sqrt{4.61} \sigma)$ confidence level. 
}
\label{FIG:ST}
\end{figure}
In FIG.~\ref{FIG:ST}, we show predictions on the $S$ and $T$
parameters in the THDM together with  the allowed regions from the
precision data for each confidence level. 
The SM-like Higgs boson mass is varied from $117$ GeV to $517$ GeV 
(black curve: from up to down), and 
the SM-like limit $\sin(\beta-\alpha)=1$ and $m_{H^\pm}^{}=300$ GeV are taken. 
We can see that electroweak precision data favor relatively light Higgs boson 
$m_h \lesssim 145$ GeV ($90\%$ CL). 
The degenerated mass of the heavy neutral Higgs bosons  $m_A^{} = m_H^{}$ is varied from 
$200$ GeV to $400$ GeV by the $10$ GeV step (dots: from left to right) for 
the given several values of the SM-like Higgs boson mass $m_h=117, 140, 240$ and $500$ GeV. 
The quadratic dependence on the mass difference between additional heavy
scalar bosons can be easily understood
by the approximate formula for $m_A^{} \sim m_{H^\pm}^{}$ in Eq.~\eqref{Eq:F5}. 
Therefore, the deviation of the $T$ parameter is insensitive to $M$. 

In the SM, the mass of the Higgs boson is constrained due to tree level unitarity. 
It has been studied by considering $6\times 6$ scattering matrix of two body 
scalar states ($hh,hz,zz,\omega^+\omega^-,h \omega^+,z\omega^+$) 
where each eigenvalues of scattering matrices are restricted  to be less than 
a criteria $\xi$ as $|a_0| \le \xi$\cite{Ref:LQT} where $a_0$ is the S wave amplitude matrix. 
For $\xi=1/2$, the Higgs boson mass is bounded to be less than about $710$ GeV. 
In the THDM, there are $14$ neutral \cite{Ref:KKT}, $8$ singly charged and a doubly charged 
two body states\cite{Ref:AAN}. 
In our numerical analysis, absolute values of all eigenvalues for the
S wave amplitude matrix are required  
to be less than $1/2$ as for a criteria to keep perturbativity\cite{Ref:HHG}. 
%
For the constraint from vacuum stability, the Higgs potential is
required to be positive for a large value of the order parameter.
In the SM, this condition is expressed by $\lambda>0$ at the tree level.
In the THDM, the condition for vacuum stability is given by \cite{Ref:DM,Ref:NS,Ref:KKO}
\begin{align}
\sqrt{\lambda_1\lambda_2}+\lambda_3
+\min\left[0,\lambda_4-\left|\lambda_5\right|\right]>0.
\end{align}

\begin{figure*}[tb]
\centering
\includegraphics[height=7.1cm]{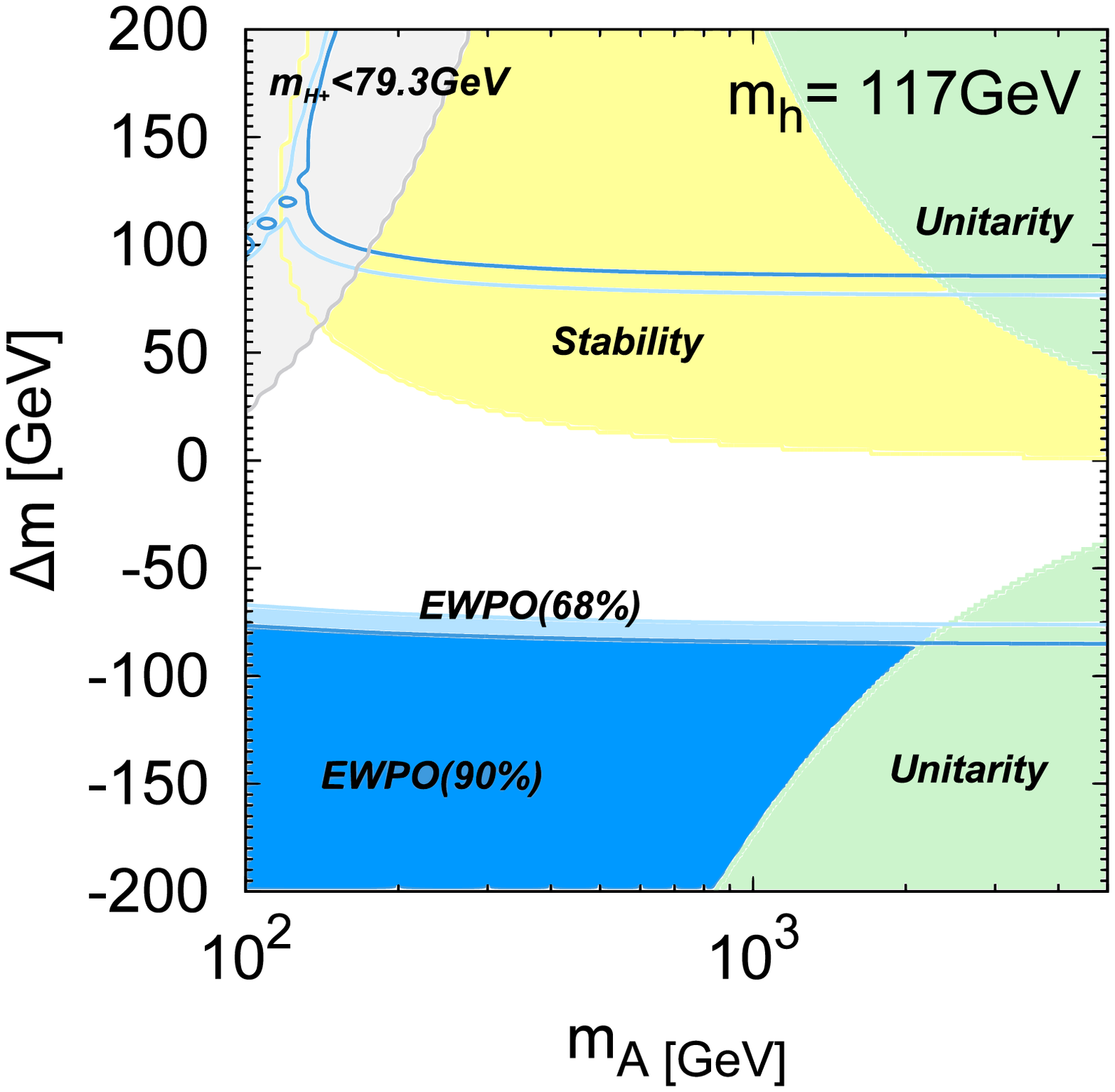}
\includegraphics[height=7.0cm]{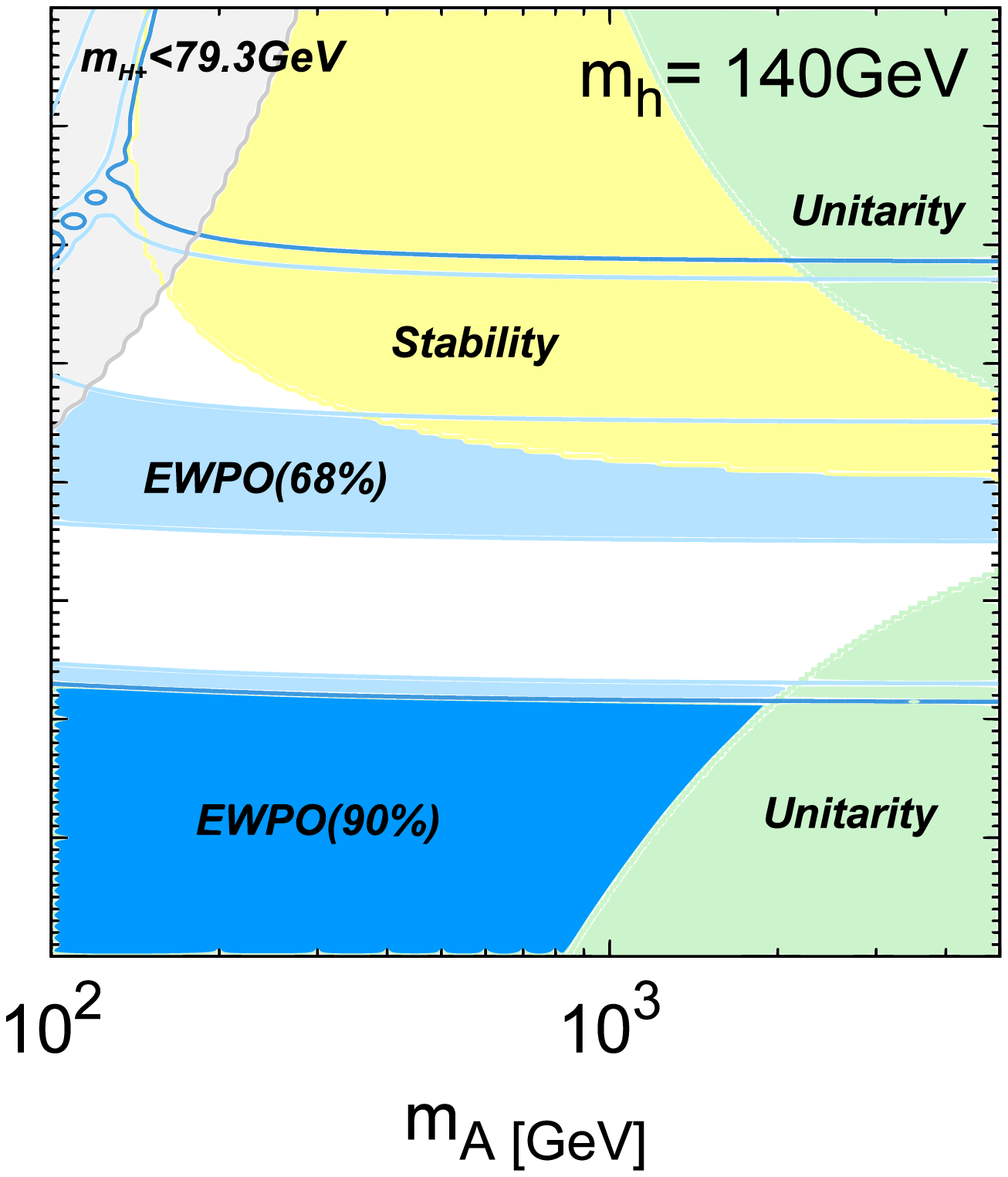}
\includegraphics[height=7.1cm]{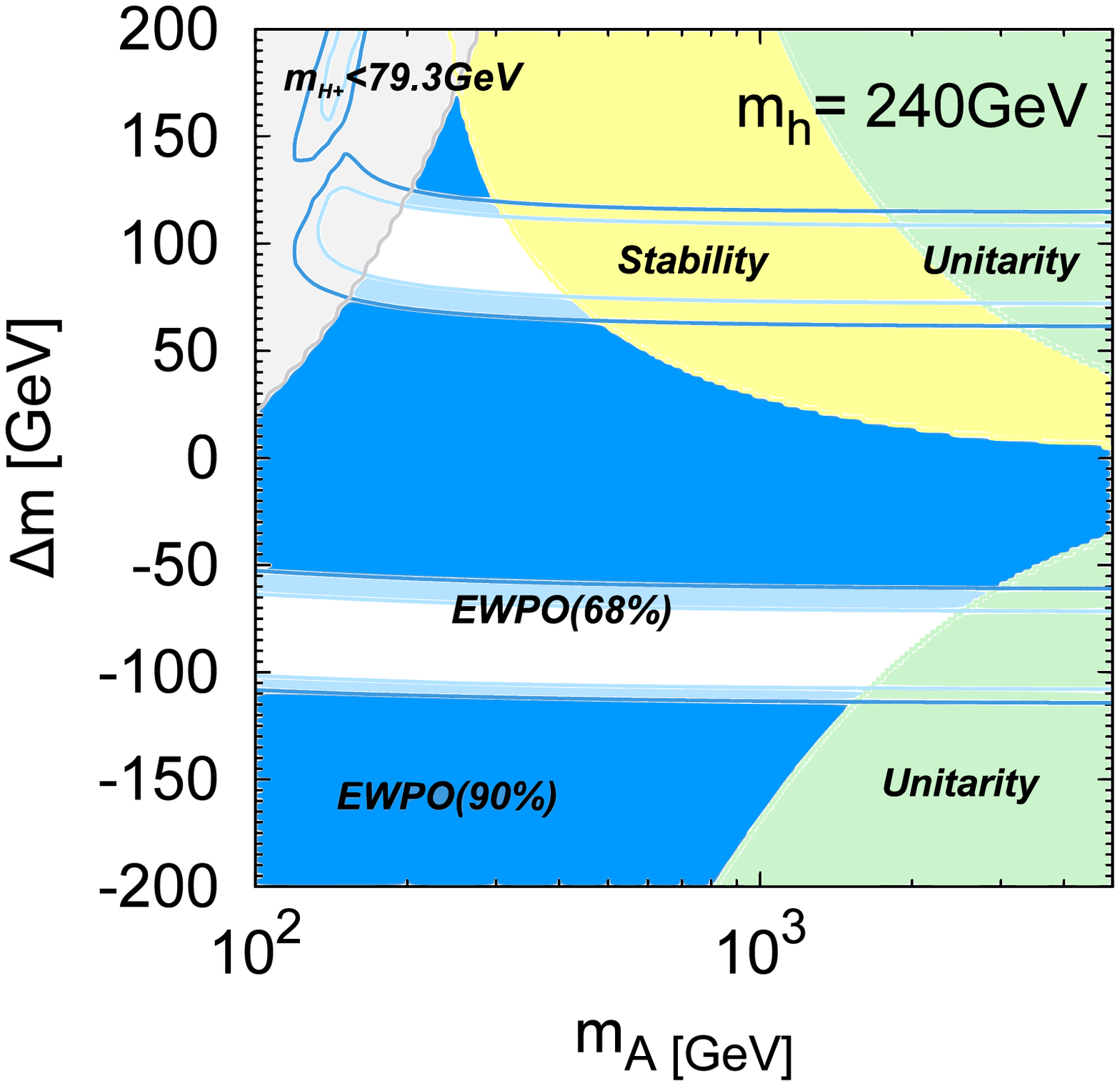}
\includegraphics[height=7.0cm]{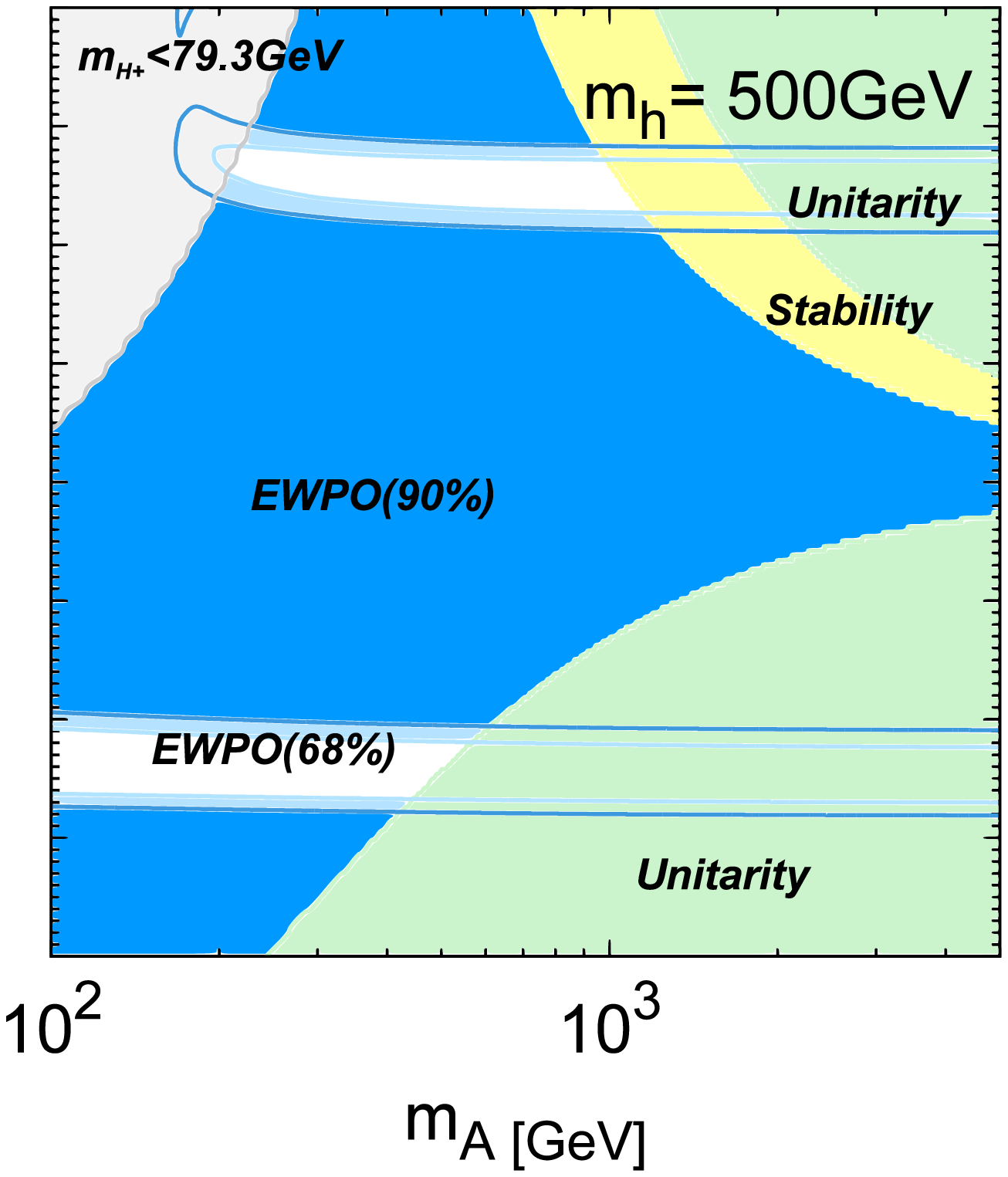}
\caption{Theoretical and experimental constraints in the parameter 
space of the THDM. 
Uncolored regions are allowed by all the constraints we here considered, 
i.e., tree level unitarity/stability and electroweak precision data, 
and direct search bound of charged Higgs boson, $m_{H^\pm} < 79.3$ GeV. 
The mass and mixing parameters are chosen as $M^2=m_H^2=m_A^2$, with the SM-like limit 
$\sin(\beta-\alpha)=1$. In this limit, constraints are independent from $\tan\beta$.}
\label{FIG:KOTT}
\end{figure*}
In FIG.~\ref{FIG:KOTT}, we show the regions excluded by various theoretical 
and experimental constraints in the THDM parameter space
(on the $m_A^{}$--$\Delta m$ plane) assuming 
the SM-like Higgs boson mass to be $m_h=117, 140, 240$ and $500$ GeV 
with $\sin(\beta-\alpha)=1$, where $\Delta m =m_A^{}-m_{H^\pm}^{}$. 
The masses of neutral scalars and the soft-breaking mass parameter
are taken to be degenerate $m_A^2=m_H^2=M^2$. 
Since quartic coupling constants $\lambda_i$ are independent on $\tan\beta$
for $m_H^{} = M$ with the SM-like limit, unitarity and stability 
bounds do not depend on $\tan\beta$ in the same limit. 
Regions excluded by the conditions from tree level unitarity and vacuum
stability are shown as the green and yellow areas, respectively,
while that excluded by the electroweak precision data at the $90\%$ 
$(68\%)$ CL is indicated by the blue (light-blue) area. 
Although the bounds from perturbative unitarity, vacuum stability and
the electroweak precision data with the oblique corrections do not
depend on the type of Yukawa interaction, the direct search results
for the charged Higgs boson depends on that via the decay process. 
The region with charged Higgs boson mass below $79.3$ GeV\cite{Ref:PDG} 
is shown as the gray area, which is excluded assuming 
${\mathcal B}(H^+\to\tau^+ \nu)+{\mathcal B}(H^+\to c \bar s)$=1. 
Depending on the type of Yukawa interaction~\cite{Ref:Barger,Ref:AKTY}, 
we may have additional constraints from the flavor physics data analyses
such as $B_s\to X_s\gamma$~\cite{Ref:Barger,bsg,bsg2}, $B^+ \to \tau^+
\nu$~\cite{btaunu} and $\tau \to \ell \nu \bar\nu (\ell =
e,\mu)$~\cite{tauleptonicdecay,tauleptonicdecay2}. 
We do not consider these constraints in FIG.~\ref{FIG:KOTT} 
because they are model-dependent.
In the upper two panels for $m_h=117$ GeV and $m_h=140$ GeV, entire 
regions of $m_A^{}$ ($< 5$ TeV) are allowed by all the constraints for
a relatively small value for $|\Delta m|$. 
On the other hand, in the lower two panels for $m_h=240$ GeV and $m_h=500$ 
GeV, we can see that deviations from the allowed region by the
electroweak precision data 
require new contributions to the electroweak precision parameters; i.e.,
a relatively large value of $|\Delta m|$.
Since quartic coupling constants are constrained by 
the tree level unitarity, masses of heavy scalar bosons 
are essentially determined by the magnitude of $M^2$ for
$M^2 \gg v^2$, where the mass difference $\Delta m$ is expressed by 
$\Delta m \simeq (\lambda_i-\lambda_j)v^2/M$. 
For a given $m_h$, the magnitude of $\Delta m$ is determined to
satisfy the electroweak precision data via the new $T$ parameter
contribution, which is proportional to $(\Delta m)^2$.
Consequently, $m_A^{}$ is constrained from above by unitarity bounds.
For $m_h=240$ GeV and $m_h=500$ GeV cases, 
we have the upper bound to be $2$ TeV and $1$ TeV, respectively. 

We have shown the results in the case with $\sin(\beta-\alpha)=1$,  
and $m_H^2=m_A^2=M^2$ in FIG.~\ref{FIG:KOTT}.
This choice of the parameters would be rather special
in the sense that there is no $\tan\beta$ dependence in this case. 
In Eqs.~\eqref{Eq:lam1} and \eqref{Eq:lam2} with $\sin(\beta-\alpha)=1$, 
terms dependent on $\tan\beta$ are proportional to $m_H^2-M^2$, and then 
one of $\lambda_1$ and $\lambda_2$ tends to large when $\tan\beta\ne 1$. 
Therefore, the parameter space is more restricted by the unitarity constraints 
in the case without degeneracy. 
Also when $\sin(\beta-\alpha)=1$ is slightly relaxed, 
the bound from tree level unitarity becomes sensitive to $\tan\beta$. 
For larger values of $\tan\beta$, the bound becomes more restrictive. 
Consequently, the tree level unitarity bound
shown in FIG.~\ref{FIG:KOTT} can be regarded as the most conservative
which is independent of the values of $\tan\beta$. 

Finally, we shortly discuss the implication to the collider phenomenology.
At the LHC, a heavy SM-like Higgs boson can be found via 
the gluon fusion $gg \to h$ or vector boson fusion 
$VV \to h$ ($V=W$ and $Z$) with the decays 
into $WW^{(\ast)}$ and $ZZ^{(\ast)}$ \cite{SM-Higgs}. 
The additional neutral scalar bosons $\phi$ ($=H$ and $A$) 
would be produced via gluon fusion 
$gg \to \phi$~\cite{gf}, associated production with heavy quarks
$pp \to t \bar t \phi$, $b \bar b \phi$~\cite{ffH},
pair production $pp \to W^{\pm\ast} \to \phi H^\pm$~\cite{AH+,AH0,Ref:AKTY} and
$pp \to Z^\ast \to AH$~\cite{AH0,Ref:AKTY}, charged Higgs boson production would be via 
$g b \to H^\pm t$~\cite{gbH+}, $pp \to W^\pm H^\mp$~\cite{WH+} and 
$pp \to H^+ H^-$~\cite{H+H-}.  
As we discussed above, if the SM-like Higgs boson is heavy,
a large mass splitting between additional heavy Higgs bosons 
$H^\pm$ and $A$ (or $H$) is required.  
In such a case, their decays into a lighter scalar boson associated with
a weak gauge boson $H^\pm \to \phi W^\pm$ 
(or $\phi \to H^\pm W^\mp$) can be significant~\cite{Akeroyd}.
These decay modes are kinematically suppressed by the degeneracy of 
scalar boson masses in most of previous discussions, for example, on Higgs boson decays
in the minimal supersymmetric SM\cite{djouadi}.  
In addition, the one-loop induced decay process $H^\pm \to W^\pm Z$ can also be enhanced 
when the mass difference between $A$ and $H^\pm$ is large~\cite{WZ}. 
The bosonic decay branching fractions of scalar bosons can dominate 
over their fermionic decay modes. 
Therefore, detailed studies for these decay modes
will be important to test the scenario with large mass splitting between
additional heavy Higgs bosons~\cite{general}.

In conclusion, we have analyzed theoretical bounds and experimental constraints 
in the THDM. For a given Higgs boson mass, the magnitude of the mass
difference between additional heavy scalar bosons
can be determined to satisfy the electroweak precision data.
However, the mass difference requires a large coupling constant in the
  Higgs potential, and too large coupling constant violates 
  tree level unitarity. 
Therefore, we have found that an upper bound on the additional heavy
Higgs bosons is obtained when the SM Higgs boson is heavy.
For example, $m_A^{}$ is bounded to be less than around $2$ $(1)$ TeV for
$m_h=240$ $(500)$ GeV,
where $M^2=m_H^2=m_A^2$ with the SM-like limit $\sin(\beta-\alpha)=1$ is taken. 
Even if the SM-like Higgs boson is found to be light $(\lesssim 140$ GeV) 
our analysis show a possible range of mass splitting in the heavy Higgs bosons 
in the THDM. \\

\noindent {\bf Acknowledgment}\\
The research of Y.O. is supported in part by the
Grant-in-Aid for Science Research, Japan Society for the Promotion of
Science (JSPS), No. 20244037 and No. 22244031. 
The research of S.K. is supported in part by the
Grant-in-Aid for Science Research, Japan Society for the Promotion of
Science (JSPS), No. 22244031. 
The work of K.T. is supported in part by the National Science Council of Taiwan
under Grant No. NSC 99-2811-M-002-088. 


\begin{thebibliography}{99}
\bibitem{Ref:PDG}
  K.~Nakamura {\it et al.}  [Particle Data Group],
  J.\ Phys.\ G {\bf 37} (2010) 075021.

\bibitem{Ref:LEP-h}
  R.~Barate {\it et al.} [ LEP Working Group for Higgs boson searches and ALEPH and DELPHI and L3 and OPAL Collaborations ],
  Phys.\ Lett.\  {\bf B565 } (2003)  61-75.

\bibitem{Ref:Tevatron-h}
 The CDF, D0 Collaborations, the Tevatron New Phenomena, Higgs Working Group, 
  arXiv:1107.5518 [hep-ex]. 

\bibitem{Ref:atlas-h}
ATLAS-CONF-2011-112.

\bibitem{Ref:lhc-h}
Plenary talk by E.~James at EPS2011, 
\url{http://indico.in2p3.fr/getFile.py/access?contribId=984&sessionId=16&resId=0&materialId=slides&confId=5116}

\bibitem{Ref:LQT}
  B.~W.~Lee, C.~Quigg, H.~B.~Thacker,
  Phys.\ Rev.\ Lett.\  {\bf 38 } (1977)  883-885;
%
  Phys.\ Rev.\  {\bf D16 } (1977)  1519.

\bibitem{Ref:DW}
  S.~Dawson, S.~Willenbrock,
  Phys.\ Rev.\ Lett.\  {\bf 62 } (1989)  1232.

\bibitem{Ref:WW}
  M.~S.~Chanowitz, M.~K.~Gaillard,
  Phys.\ Lett.\  {\bf B142 } (1984)  85;
%
  Nucl.\ Phys.\  {\bf B261 } (1985)  379.

\bibitem{Ref:Triv}
  M.~Lindner,
  Z.\ Phys.\  {\bf C31 } (1986)  295;
%
  B.~Grzadkowski, M.~Lindner,
  Phys.\ Lett.\  {\bf 178B } (1986)  81.

\bibitem{Ref:HR}
  T.~Hambye, K.~Riesselmann,
  Phys.\ Rev.\  {\bf D55 } (1997)  7255-7262.

\bibitem{Ref:KM}
  C.~F.~Kolda, H.~Murayama,
  JHEP {\bf 0007 } (2000)  035.

\bibitem{Ref:PeskinTakeuchi}
  M.~E.~Peskin, T.~Takeuchi,
  Phys.\ Rev.\ Lett.\  {\bf 65 } (1990)  964-967; 
%
  Phys.\ Rev.\  {\bf D46 } (1992)  381-409.

\bibitem{Ref:LEP-EW}
 LEP and ALEPH and DELPHI and L3 and LEP Electroweak Working Group and SLD Electroweak Group and SLD Heavy Flavour Group and OPAL Collaborations,
  [hep-ex/0412015].

\bibitem{Ref:ST-2hdm}
  D.~Toussaint,
  Phys.\ Rev.\  {\bf D18 } (1978)  1626.

\bibitem{STUinTHDM}
  S.~Bertolini,
  Nucl.\ Phys.\  {\bf B272 } (1986)  77.

\bibitem{Ref:KKT}
  S.~Kanemura, T.~Kubota, E.~Takasugi,
  Phys.\ Lett.\  {\bf B313 } (1993)  155-160.

\bibitem{Ref:AAN}
  A.~G.~Akeroyd, A.~Arhrib, E.~-M.~Naimi,
  Phys.\ Lett.\  {\bf B490 } (2000)  119-124.

\bibitem{Ref:DM}
  N.~G.~Deshpande, E.~Ma,
  Phys.\ Rev.\  {\bf D18 } (1978)  2574.

\bibitem{Ref:NS}
S.~Nie, M.~Sher,
  Phys.\ Lett.\  {\bf B449 } (1999)  89-92.
Z

\bibitem{Ref:KKO}

  S.~Kanemura, T.~Kasai, Y.~Okada,
  Phys.\ Lett.\  {\bf B471 } (1999)  182-190.

\bibitem{Ref:Barger}
  V.~D.~Barger, J.~L.~Hewett, R.~J.~N.~Phillips,
  Phys.\ Rev.\  {\bf D41 } (1990)  3421;
%
  Y.~Grossman,
  Nucl.\ Phys.\  {\bf B426 } (1994)  355-384.

\bibitem{bsg}
  M.~Ciuchini, E.~Franco, G.~Martinelli, L.~Reina, L.~Silvestrini,
  Phys.\ Lett.\  {\bf B334 } (1994)  137-144;
%
  M.~Ciuchini, G.~Degrassi, P.~Gambino, G.~F.~Giudice,
  Nucl.\ Phys.\  {\bf B527 } (1998)  21-43;
%
  F.~Borzumati, C.~Greub,
  Phys.\ Rev.\  {\bf D58 } (1998)  074004;
%
  P.~Gambino, M.~Misiak,
  Nucl.\ Phys.\  {\bf B611 } (2001)  338-366.

\bibitem{btaunu}
  W.~-S.~Hou,
  Phys.\ Rev.\  {\bf D48 } (1993)  2342-2344;
%
  Y.~Grossman, Z.~Ligeti,
  Phys.\ Lett.\  {\bf B332 } (1994)  373-380;
%
  Y.~Grossman, H.~E.~Haber, Y.~Nir,
  Phys.\ Lett.\  {\bf B357 } (1995)  630-636.

\bibitem{tauleptonicdecay}
  W.~Hollik, T.~Sack,
  Phys.\ Lett.\  {\bf B284 } (1992)  427-430.

\bibitem{tauleptonicdecay2}
 M.~Krawczyk, D.~Temes,
  Eur.\ Phys.\ J.\  {\bf C44 } (2005)  435-446.

\bibitem{bsg2}
 M.~Misiak, H.~M.~Asatrian, K.~Bieri, M.~Czakon, A.~Czarnecki, T.~Ewerth, A.~Ferroglia, P.~Gambino {\it et al.},
  Phys.\ Rev.\ Lett.\  {\bf 98 } (2007)  022002.

\bibitem{Ref:GW}
  S.~L.~Glashow, S.~Weinberg,
  Phys.\ Rev.\  {\bf D15 } (1977)  1958.

\bibitem{kosy}
  S.~Kanemura, Y.~Okada, E.~Senaha and C.~P.~Yuan,
  Phys.\ Rev.\  D {\bf 70} (2004) 115002. 

\bibitem{pomarol}
  H.~E.~Haber and A.~Pomarol,
  Phys.\ Lett.\  B {\bf 302} (1993) 435;
%
  A.~Pomarol and R.~Vega,
  Nucl.\ Phys.\  B {\bf 413} (1994) 3.

\bibitem{Ref:HHG}
  J.~F.~Gunion, H.~E.~Haber, G.~Kane and S.~Dawson,
  The Higgs Hunter's Guide
  (Frontiers in Physics series, Addison-Wesley, Reading, MA, 1990).

\bibitem{Ref:AKTY}
  M.~Aoki, S.~Kanemura, K.~Tsumura, K.~Yagyu,
  Phys.\ Rev.\  {\bf D80 } (2009)  015017.

\bibitem{SM-Higgs}
  For a review, see:  
%
  A.~Djouadi,
  Phys.\ Rept.\  {\bf 457} (2008) 1.

\bibitem{gf}
  H.~M.~Georgi, S.~L.~Glashow, M.~E.~Machacek and D.~V.~Nanopoulos,
  Phys.\ Rev.\ Lett.\  {\bf 40} (1978) 692.

\bibitem{ffH}
  S.~Dawson, C.~B.~Jackson, L.~Reina and D.~Wackeroth,
  Phys.\ Rev.\  D {\bf 69} (2004) 074027;
%
  S.~Dittmaier, M.~1.~Kramer and M.~Spira,
  Phys.\ Rev.\  D {\bf 70} (2004) 074010;
%
  J.~F.~Gunion, H.~E.~Haber, F.~E.~Paige, W.~K.~Tung and S.~S.~D.~Willenbrock,
  Nucl.\ Phys.\  B {\bf 294} (1987) 621;
%
  Z.~Kunszt,
  Nucl.\ Phys.\  B {\bf 247} (1984) 339.
 
\bibitem{AH+}
  S.~Kanemura and C.~P.~Yuan,
  Phys.\ Lett.\  B {\bf 530} (2002) 188;
%
  A.~G.~Akeroyd,
  Phys.\ Rev.\  D {\bf 68} (2003) 077701; 
%
  Q.~H.~Cao, S.~Kanemura and C.~P.~Yuan,
  Phys.\ Rev.\  D {\bf 69} (2004) 075008;
%
  A.~Belyaev, Q.~H.~Cao, D.~Nomura, K.~Tobe and C.~P.~Yuan,
  Phys.\ Rev.\ Lett.\  {\bf 100} (2008) 061801.
        
\bibitem{AH0}
  J.~F.~Gunion and H.~E.~Haber,
  Nucl.\ Phys.\  B {\bf 278} (1986) 449.
        
\bibitem{gbH+}
  T.~Plehn,
  Phys.\ Rev.\  D {\bf 67} (2003) 014018; 
%
  E.~L.~Berger, T.~Han, J.~Jiang and T.~Plehn,
  Phys.\ Rev.\  D {\bf 71} (2005) 115012.
        
\bibitem{WH+}
  A.~A.~Barrientos Bendezu and B.~A.~Kniehl,
  Phys.\ Rev.\  D {\bf 59} (1999) 015009;
%
  O.~Brein, W.~Hollik and S.~Kanemura,
  Phys.\ Rev.\  D {\bf 63} (2001) 095001; 
%
  E.~Asakawa, O.~Brein and S.~Kanemura,
  Phys.\ Rev.\  D {\bf 72} (2005) 055017; 
%
  D.~Eriksson, S.~Hesselbach and J.~Rathsman,
  Eur.\ Phys.\ J.\  C {\bf 53} (2008) 267.  

\bibitem{H+H-}
  A.~Alves and T.~Plehn,
  Phys.\ Rev.\  D {\bf 71} (2005) 115014; 
%
  S.~Moretti,
  J.\ Phys.\ G {\bf 28} (2002) 2567.

\bibitem{Akeroyd}
  A.~G.~Akeroyd,
  Nucl.\ Phys.\  B {\bf 544} (1999) 557; 
%
  A.~G.~Akeroyd, A.~Arhrib, E.~Naimi,
  Eur.\ Phys.\ J.\  {\bf C20 } (2001)  51-62; 
%
  F.~Borzumati and A.~Djouadi,
  Phys.\ Lett.\  B {\bf 549} (2002) 170.

\bibitem{djouadi}        
  A.~Djouadi,
  Phys.\ Rept.\  {\bf 459} (2008) 1.

\bibitem{WZ}
  S.~Kanemura,
  Phys.\ Rev.\  {\bf D61 } (2000)  095001.

\bibitem{general}
  S.~Kanemura, S.~Moretti, Y.~Mukai, R.~Santos, K.~Yagyu,
  Phys.\ Rev.\  {\bf D79 } (2009)  055017.
%
  G.~C.~Branco, P.~M.~Ferreira, L.~Lavoura, M.~N.~Rebelo, M.~Sher and J.~P.~Silva,
  arXiv:1106.0034 [hep-ph].
\end{thebibliography}
\end{document}